\documentclass[letterpaper]{article} 
\usepackage[submission]{aaai25}  
\usepackage{times}  
\usepackage{helvet} 
\usepackage{courier}  
\usepackage[hyphens]{url}  
\usepackage{graphicx} 
\usepackage{natbib}  
\usepackage{caption} 
\usepackage{algorithm}
\usepackage{algorithmic}

\usepackage{newfloat}
\usepackage{listings}
\DeclareCaptionStyle{ruled}{labelfont=normalfont,labelsep=colon,strut=off} 
\floatstyle{ruled}
\newfloat{listing}{tb}{lst}{}
\floatname{listing}{Listing}

\title{Verification methods for international AI agreements}
\author{
   Akash R. Wasil\textsuperscript{\rm 1 \rm 2}, Tom Reed\textsuperscript{\rm 2}, Jack William Miller\textsuperscript{\rm 2} and Peter Barnett\textsuperscript{\rm 3}
}
\affiliations{
    \textsuperscript{\rm 1}Georgetown University\\
    \textsuperscript{\rm 2} University of Cambridge, ERA AI Fellowship \\
    \textsuperscript{\rm 3} Independent
}

\usepackage{xcolor}
\usepackage{tcolorbox}
\usepackage{enumitem}

\usepackage{xcolor}

\usepackage{longtable}
\usepackage{array} 

\usepackage{cleveref}
\crefname{figure}{Figure}{Figures}

\begin{document}

\onecolumn

\vspace*{\fill}

\begin{center}
    {\huge \textbf{Verification methods for international AI agreements}}
    \vspace{1em}

    \textbf{Akash R. Wasil}$^{1,2}$, 
    \textbf{Tom Reed}$^{2}$, 
    \textbf{Jack William Miller}$^{2}$, 
    \textbf{Peter Barnett}$^{3}$ \\\vspace{0.25em}
    
    $^{1}$Georgetown University (\texttt{aw1404@georgetown.edu}) \\\vspace{0.25em}
    
    $^{2}$University of Cambridge, ERA AI Fellowship \\\vspace{0.25em}
    
    $^{3}$Independent \\

    \vspace{2em}
\end{center}

\begin{abstract}
\centering
\begin{minipage}[t]{0.8\textwidth}
\Large

What techniques can be used to verify compliance with international agreements about advanced AI development? In this paper, we examine 10 verification methods that could detect two types of potential violations: unauthorized AI training (e.g., training runs above a certain FLOP threshold) and unauthorized data centers. We divide the verification methods into three categories: (a) national technical means (methods requiring minimal or no access from suspected non-compliant nations), (b) access-dependent methods (methods that require approval from the nation suspected of unauthorized activities), and (c) hardware-dependent methods (methods that require rules around advanced hardware). For each verification method, we provide a description, historical precedents, and possible evasion techniques. We conclude by offering recommendations for future work related to the verification and enforcement of international AI governance agreements.
\end{minipage}
\end{abstract}

\vspace*{\fill}

\newpage

\vspace{1em}

\section{\huge Executive Summary} 

\vspace{2em}

{
\setlength{\parindent}{0pt}
\large
Efforts to maximize the benefits and minimize the global security risks of advanced AI may lead to international agreements. This paper outlines methods that could be used to verify compliance with such agreements. The verification methods we cover are focused on detecting two potential violations: 
\vspace{0.5em}
\begin{tcolorbox}[title=Violations to verify]
\begin{itemize}
\item \textbf{Unauthorized AI development} (for example, AI development that goes beyond a FLOP threshold set by an international agreement, or the execution of a training run that has not received a license).
\item  \textbf{Unauthorized data centers} (for example, data centers that go beyond a maximum computing capacity limit or networking limit set by an international agreement). 
\end{itemize}
\end{tcolorbox}
\vspace{0.5em}
We identify \textbf{10 verification methods} and divide them into three categories: 
\begin{enumerate}
\item \textbf{National technical means.} Methods that can be used by nations unilaterally.
\item \textbf{Access-dependent methods.} Methods that require a nation to grant access to national or international inspectors
\item \textbf{Hardware-dependent methods.} Methods that require agreements pertaining to advanced hardware 
\end{enumerate}
\vspace{0.5em}
\begin{tcolorbox}[title=National technical means]
\begin{enumerate}
\item \textbf{Remote sensing}: Detect unauthorized data centers and semiconductor manufacturing via visual and thermal signatures.
\item \textbf{Whistleblowers}: Incentivize insiders to report non-compliance.
\item \textbf{Energy monitoring}: Detect power consumption patterns that suggest the potential presence of large GPU clusters.
\item \textbf{Customs data analysis}: Track the movement of critical AI hardware and raw materials. 
\item \textbf{Financial intelligence}: Monitor large financial transactions related to AI development. 
\end{enumerate}
\end{tcolorbox}
\vspace{0.5em}
\begin{tcolorbox}[title=Access-dependent methods]
\begin{enumerate}
\item \textbf{Data center inspections}: Conduct inspections of sites to assess the size of a data center, verify compliance with hardware agreements, and verify compliance with other safety and security agreements.
\item \textbf{Semiconductor manufacturing facility inspections}: Conduct inspections of sites to determine the quantity of chip production and verify that chip production conforms to any agreements around advanced hardware.
\item \textbf{AI developer inspections}: Conduct inspections of AI development facilities via interviews, document and training transcript audits, and potential code reviews.
\end{enumerate}
\end{tcolorbox}
\vspace{0.5em}
\begin{tcolorbox}[title=Hardware-dependent methods]
\begin{enumerate}
\item \textbf{Chip location tracking}: Automatic location tracking of advanced AI chips.
\item \textbf{Chip-based reporting}: Automatic notification if chips are used for unauthorized purposes.
\end{enumerate}
\end{tcolorbox}
\vspace{0.5em}

\subsection{\LARGE Limitations and considerations}
The verification methods we propose have some limitations, and there are many complicated national and international considerations that would influence if and how they are implemented. Some of these include:
\begin{itemize}
\item \textbf{Invasiveness}: Some methods (especially on-site inspections) may be seen as intrusive and could raise concerns about privacy and sovereignty. Several factors could influence a nation's willingness to accept invasive measures (e.g., the amount of international tension or distrust between nations, the degree to which nations are concerned about risks from advanced AI, the exact types of risks that nations find most concerning.)

\item \textbf{Imperfect detection}: No single method is foolproof. However, the combination of multiple methods could create a ``Swiss chees'' model, where the weaknesses of one method are covered by the strengths of others.

\item \textbf{Developmental stage}: Some methods (especially the hardware-dependent ones) may require additional R\&D. Furthermore, unlike methods that have been used for decades in other areas, the real-world effectiveness of some hardware-dependent methods has not yet been determined.

\end{itemize}

\subsection{\LARGE Future Directions}
Our work provides a foundation for discussions on AI governance verification, but several key areas require further research:
\begin{itemize}
\item \textbf{Red-teaming exercises for verification regimes}. Future work could examine how adversaries might attempt to circumvent a verification regime, describe potential evasion methods, and develop robust countermeasures to improve the effectiveness of the verification regime.
\item \textbf{Design of international AI governance institutions.} Future work could examine how international AI governance institutions should be designed, potentially drawing lessons from existing international bodies. Such work could explore questions such as: (a) what specific powers should be granted to the international institution, (b) how the institution should make core decisions, (c) how power is distributed between nations, and (d) how to handle potential violations or instances of non-compliance.
\item \textbf{Enforcement strategies.} Future work could examine what kinds of responses could be issued if non-compliance is discovered. This includes examining how such responses can be proportionate to the severity of the violation. 
\item \textbf{Development of tamper-proof and privacy-preserving hardware-enabled verification mechanisms}. Future R\&D efforts could improve the effectiveness, feasibility, robustness, or desirability of various hardware-dependent verification methods. 
\end{itemize}

}

\twocolumn

\section{Introduction}
`
\begin{figure*}
    \centering
    \includegraphics[width=\textwidth]{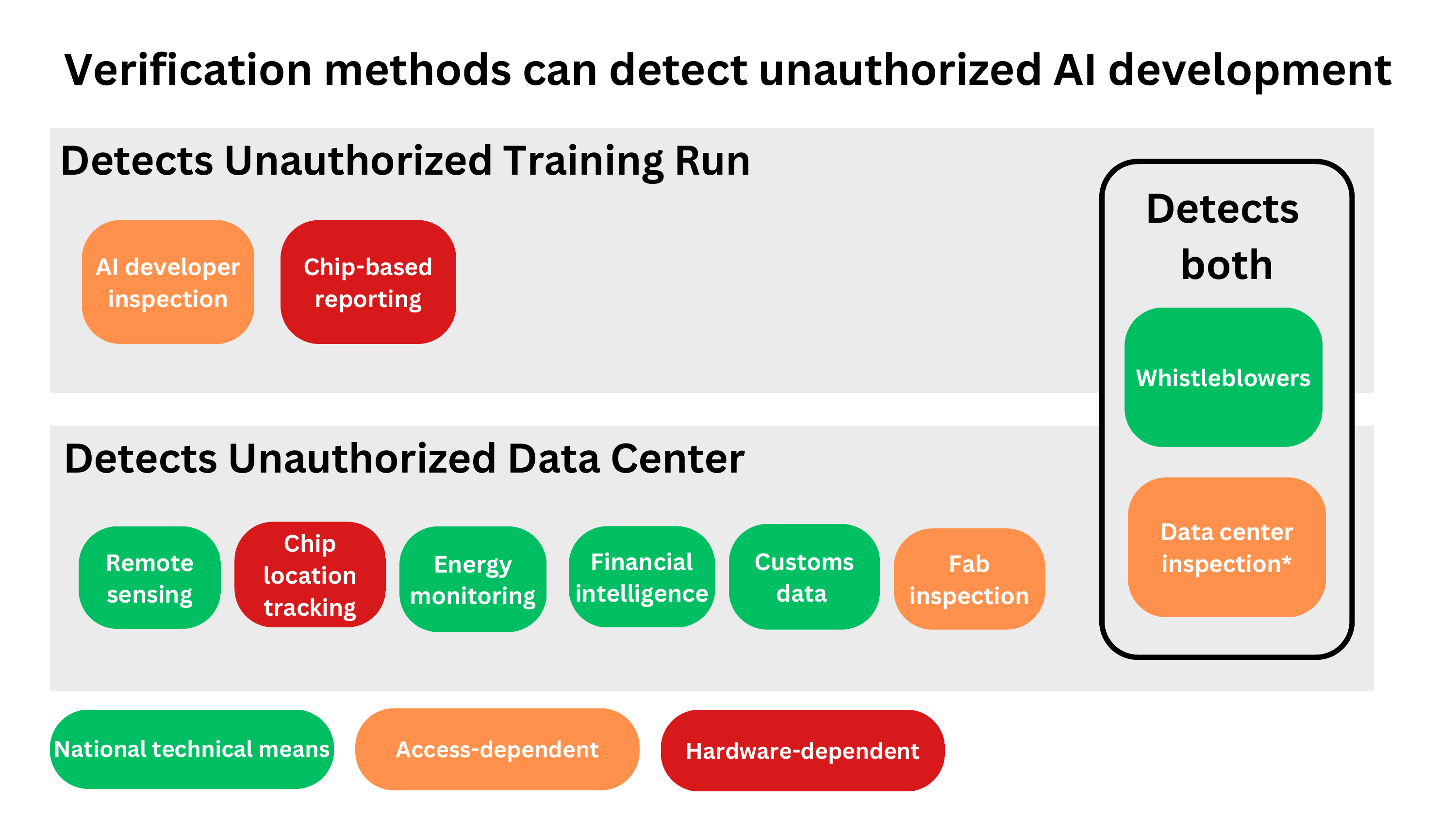}
    \caption{\small Verification methods can help detect unauthorized training runs and unauthorized data centers. * For data center inspections to be able to detect unauthorised training runs, it is likely that hardware requirements around chips with activity logs will be needed in some form.}
    \label{fig:verification-methods}
\end{figure*}

The development of advanced artificial intelligence poses major global security risks. Significant threats include the potential for pervasive surveillance, the development of autonomous weapons, and misuse by malicious actors. Some of the most concerning risks stem from loss of control and misalignment \citep{bengio2024managing, superintelligence-book, ngo2022alignment}. A sufficiently powerful misaligned AI system could autonomously act against human interests following an objective function which does not capture human values \citep{pan2022reward-mispecification}. There is a great amount of uncertainty around what kinds of safeguards will be necessary to prevent misalignment \citep{ai-values-and-alignment}. Many experts believe that safeguards may require many years or decades of concerted research effort.

AI risks are exacerbated by race dynamics --- companies are rapidly progressing in the hope of being the first to develop artificial superintelligence \citep{armstrong2016racing, hogarth2023god}. In the context of an AI race, nations may not have sufficient time to carefully and cautiously develop or evaluate such safeguards.

International agreements could help avoid or mitigate a race between nations. Even though governments are at early stages of understanding AI risks, key figures in the United States and China have already acknowledged concerns about AI global security risks and expressed interest in global governance approaches \cite{wasildurgin}. As governments become more aware of AI risks, they may become interested in global governance strategies that curb these race dynamics. Alternatively, nations might agree to cede the development of advanced AI to a joint international project. The international institution would be responsible for carrying out certain forms of advanced AI development, which would be illegal outside the context of this secure joint project \citep{hogarth2023god}.

International agreements require verification. Nations might be much more likely to form international agreements around rules that they can reliably verify \citep{fearon1995rationalist, baker2023nucleararmscontrolverification}. By ``verify'', we mean that nations would be able to detect non-compliance with agreements. Ideally, verification methods (methods used to detect non-compliance) would provide \textbf{early and reliable warning signs}. “Early”, in that non-compliance could be detected relatively quickly (before a nation achieved any meaningful unauthorized advantage in advanced AI development), and ``reliable" in that non-compliance would be very likely to be detected.\footnote{Early and reliable warning signs have also been discussed in the context of international agreements for nuclear security (see \citet{acheson1946international}, specifically pages 15 to 34.}

In this paper, we provide an overview of verification methods for international AI agreements. We begin by outlining the potential targets such an international agreement. We then outline 10 verification methods. For each verification method, we provide a description, some precedents for how the verification method has been used in the past, and an example evasion technique to illustrate how an adversary could attempt to circumvent the method. Finally, we discuss limitations of the verification methods and directions for future work.\footnote{Readers should also note that there is a forthcoming report by authors at the RAND Corporation that aims to provide a detailed examination of verification methods, analyze trade-offs with various methods, and discuss their technical implementation.}

\section{What to verify: Unauthorized AI development and unauthorized data centers}

An international agreement on AI could take many forms, depending on how the technology and its associated risks evolve. In scenarios where continued AI development leads to substantial acknowledged global security risks, we anticipate that verification methods would need to be capable of detecting two primary types of potential violations: 

\begin{enumerate}
    \item\textbf{Unauthorized data centers}. International governance of AI could plausibly set restrictions on the form, size, quantity, and location of large-scale computing facilities. Verification methods would therefore be needed to detect the construction or operation of data centers that violate these agreed-upon standards.

    \item\textbf{Unauthorized training runs}. An effective international system for governing AI would likely include restrictions on the scale and characteristics of AI development. Beyond detecting unauthorised data centers, methods to verify that \textit{known} data centers are compliant with agreed-upon standards would also be necessary. For example, an agreement might stipulate that AI training runs should not exceed a certain FLOP\footnote{FLOP stands for “floating-point operation” and reflects the amount of computations performed to train an AI system. Although no single metric is perfect, FLOP thresholds can provide a relatively straightforward way to set thresholds for advanced or dangerous AI development. As of 2024, the most powerful frontier AI systems were trained using approximately $10^{25}$ FLOP (see \citet{epoch2024trackinglargescaleaimodels}).} threshold \citep{heim2024training}, use specific types of training data, or employ certain training algorithms. Verification methods would be needed to detect whether AI development activities occurring within facilities violate such standards.  
\end{enumerate}

\section{Methodology}

Our process for compiling verification methods involved a few steps: (a) reviewing relevant literature on AI verification and international AI governance, (b) reviewing relevant literature on verification methods for agreements in other fields (e.g., nuclear security, biosecurity, arms control), and (c) conducting informal interviews with experts in technical AI governance. Through this process, we identified 10 verification methods. For each verification method, we examined its application in other fields to inform our description of how the method could be used in the context of AI disagreements and to inform our section about the method's precedent in other fields. We also grouped the methods into categories based on the circumstances in which they could be implemented: universally (national technical means), only in cases where a nation provides access (access-dependent), or only in cases in which nations have agreed to rules around the design of advanced hardware (hardware-dependent). Our process is \textit{not} intended to be systematic and our work should not be considered a comprehensive overview of verification methods. Rather, it is meant to serve as an initial step toward better understanding a set of specific verification methods and their limitations. 

\section{Verification methods}

\textbf{Defining ``verification method''.} In this piece, a verification method is \textbf{a method that could directly be used to detect defection or non-compliance from an agreement.} That is, we assume an adversarial setup in which one party is explicitly attempting to ``hide" unauthorized data centers or unauthorized AI training.

\textbf{Categorizing verification methods}. Some verification methods can be implemented without any buy-in from nations suspected of non-compliance, some verification methods require cooperation or authorization from the suspected nation, and some verification methods require cooperation from hardware manufacturers. These distinctions are useful for determining which verification methods might be feasible under various circumstances. 

Thus, we divide verification methods into three categories: (a) \textbf{national technical means} (methods that can be implemented without the approval of individual nations)\footnote{We borrow this term from the security literature, see \citet{us-state-salt1}.}, (b) \textbf{access-dependent verification methods} (methods that require international agreements that include the suspected nation), and (c) \textbf{hardware-dependent verification methods} (methods that require international agreements that include major hardware manufacturers). 
See \cref{fig:verification-methods} and \cref{fig:ast-table} for a visual summary of the verification methods.

\subsection{National Technical Means}

\subsubsection{REMOTE SENSING}

Remote sensing techniques, including satellite imagery and other forms of aerial observation, can detect potential undeclared data centers using visual, infrared and other electromagnetic signatures. Advanced commercial satellites, which can achieve sub-meter resolutions \citep{statista2022commercially}, could identify specialized cooling units and the movement of computing equipment. 

Infrared imaging is particularly promising for detecting concealed data centers, as GPUs and other computing hardware generate significant heat signatures \citep{dc-is-hot} that are difficult to mask. This could reveal large-scale computing facilities even when visually concealed.

Recent advancements in machine learning have further enhanced remote sensing capabilities for verification. Drawing from nuclear non-proliferation efforts \citep{Rutkowski2020}, AI-driven approaches such as supervised and unsupervised classification techniques can be applied to remotely sensed data. These methods could significantly improve the identification and monitoring of potential AI development facilities without requiring on-site access, bolstering national technical means for AI governance verification.

While remote sensing can be used without formal agreements, international commitments similar to START could enhance its effectiveness by ensuring non-interference and facilitating data exchange \citep{state2023new}.

\textbf{Precedent.} The International Atomic Energy Agency (IAEA) routinely employs satellite imagery to evaluate state-provided information about nuclear activities and to plan inspections \citep{baker2023nucleararmscontrolverification}.

Non-state actors have also demonstrated the power of commercial imagery; for example, the Open Nuclear Network used it to reveal maintenance and possible expansion at China's Lop Nur nuclear testing site, including new tunneling and drilling activities \citep{opennuclearnetwork2024strengthening}.

\subsubsection{WHISTLEBLOWERS}

Insiders with knowledge of undeclared facilities or operations could provide valuable information not detectable through external means. Potential whistleblowers include employees, contractors, or local residents aware of suspicious activities. Governments could incentivize whistleblowing by:
\begin{enumerate}
    \item Establishing robust protection frameworks specifically for AI and technology sectors;
    \item Offering financial incentives for verified information;
    \item Creating secure, anonymous reporting channels;
    \item Providing legal support and job protection;
    \item Developing international cooperation for cross-border whistleblower protection \citep{Loyens2018}.
\end{enumerate}

It is important to note that incentivization alone may not be sufficient to ensure the effectiveness of whistleblower schemes, given that determined adversaries might attempt to physically or digitally block employees from contacting a verifying authority.

One possible solution to this limitation is to implement regular in-person communication with employees, such as through semi-structured interviews \citep{wasil2024understanding}.\footnote{We thank Mauricio Baker for this suggestion, which is also included in his upcoming report.}

\textbf{Precedent.} The SEC Whistleblower Program, established under the Dodd-Frank Act in 2010, created a system for reporting securities violations \citep{sec2017whistleblower}. It includes strong protections and incentives like monetary awards for whistleblowers, who are entitled to anywhere between 10-30\% of the sanctions resulting from their information \citep{reuters2018sec}. For example, in 2016, three whistleblowers revealed Merrill Lynch's misuse of up to \$58 billion daily in customer funds, leading to a \$415 million settlement and \$83 million in whistleblower awards \citep{sec2016merrill,reuters2018sec}.

\subsubsection{ENERGY MONITORING}

Unauthorized data centers or the use of data centers for unauthorized training runs could be detected by monitoring energy consumption, either passively (through grid data obtained via espionage), or actively (using devices to measure grid activity).  

If the total amount of energy reaching a data center can be measured with reasonable accuracy, it should be possible to convert the energy estimate into a reasonable approximation of the number of FLOPs completed by that facility \citep{desislavov2021compute}. Using the FLOP/s, we could ascertain whether the facility is at an unauthorized size. However, these coarse-grained measurements may only be capable of detecting large-scale violations, and further research is needed to understand how such measurements can be more accurately translated into relevant units like FLOPs. More precise methods of energy monitoring may be necessary for detecting smaller-scale violations or unauthorized training runs.

\textbf{Precedent.} Economists use energy monitoring to verify economic data. For example, economists have used discrepancies between reported GDP growth and energy consumption to suggest exaggerated growth figures in China \citep{owyang2017china}. This principle could be applied to detect unauthorized data centers, given the direct relationship between energy consumption and FLOPS.

\subsubsection{CUSTOMS DATA ANALYSIS}

Governments can use customs data to track the movement of key components for large-scale AI computing facilities. Import and export records could be analyzed to identify unusual or unexplained patterns in the movement of critical hardware, equipment or raw materials. A sudden surge in imports of high-performance GPUs or other critical components to a specific region of concern, far exceeding the known requirements of declared facilities in that region, would indicate non-compliance. 
 
\textbf{Precedent.} The U.S. government's End-Use Monitoring (EUM) programs, particularly the Blue Lantern program for direct commercial sales, provide a robust precedent for tracking and verifying the use of sensitive technologies \citep{state2021end}. Under the Blue Lantern program, the Department of State conducts pre-license, post-license/pre-shipment, and post-shipment checks to verify the legitimacy of proposed transactions and ensure compliance with use, transfer, and security requirements. This program has been successful in promoting understanding of U.S. defense trade controls, building mutual confidence among stakeholders, mitigating risks of diversion and unauthorized use, and uncovering violations of the Arms Export Control Act. A similar approach could be adapted for monitoring the movement and use of critical AI hardware components in countries at different stages of the chip supply chain. 

\subsubsection{FINANCIAL INTELLIGENCE}

Governments could track suspicious financial transactions relating to the purchase of important components of AI development. Financial institutions could be required to flag large or unusual purchases of specialized AI hardware, monitor transactions to known AI chip manufacturers, and cross-reference financial data with customs information.

\textbf{Precedent.} In the US, the Financial Crimes Enforcement Network (FinCEN) uses the Suspicious Activity Report (SAR) system and FinCEN Exchange, a public-private partnership, to combat money laundering, terrorism financing, and organised crime \citep{fincen2024advisory}.   

In the early 2010s, an SAR filed by a bank led to the discovery of a complex international bribery scheme. The case resulted in multiple arrests and the seizure of over \$100 million in criminal proceeds \citep{fincen2011sar}. This demonstrates how financial intelligence can uncover sophisticated international financial crimes, potentially adaptable to detecting undeclared AI development activities.  

\begin{figure}[ht!]
    \centering
    \includegraphics[width=1\linewidth]{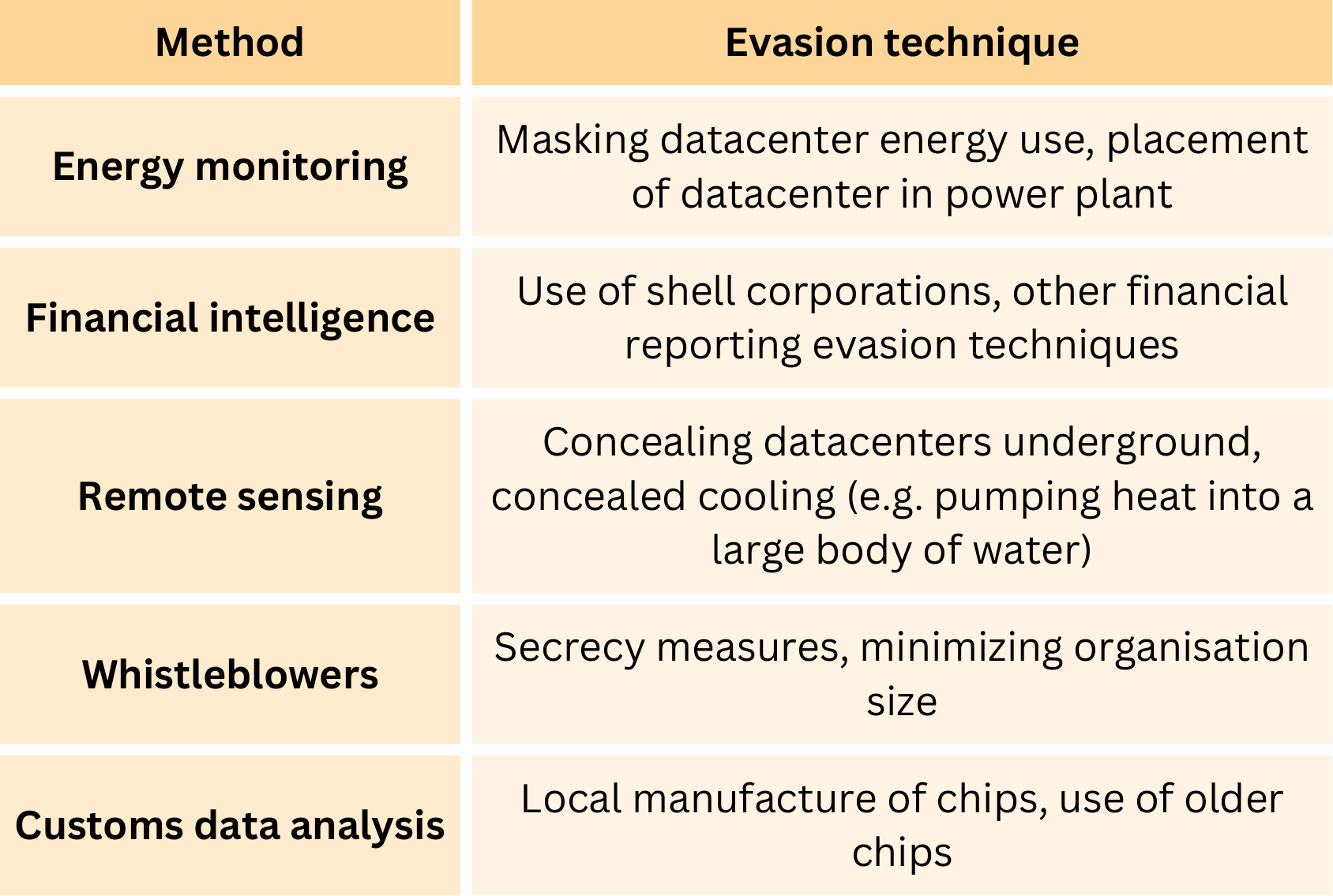}
    \caption{Summary of evasion techniques to avoid verification methods under national technical means.}
    \label{fig:national-technical-counters}
\end{figure}

\subsubsection{KEY TAKEAWAYS}

National technical means offer a valuable starting point for verifying compliance with AI governance agreements. Nations already have extensive experience using these methods to verify compliance with other kinds of international agreements. These methods can plausibly be used to detect large-scale AI infrastructure and unusual patterns in energy consumption, hardware imports, and financial transactions. However, these methods have important limitations. In particular, adversaries could attempt to disguise data centers as other high-energy facilities like power plants, or when compute is distributed across multiple smaller sites.

\subsection{Access-dependent verification methods}

\subsubsection{ON-SITE INSPECTIONS OF DATA CENTERS}

On-site inspections involve physical visits to declared data centers to verify compliance with agreements on computing power. These inspections would focus on several aspects, including (but not limited to):

\begin{itemize}
    \item \textit{Chip identifiers.} AI-capable chips could be required to have unique identifiers \citep{Aarne2024}. Inspectors could catalog these identifiers to ensure they match declared inventories. 
    \item \textit{Chip activity logs.} Require chips to have activity logs that inspectors can analyze to verify that: (1) chips are being used in accordance with their declared purposes and within agreed-upon limits, and (2) only licensed code is being executed on the chips \citep{Shavit2023}.
    \item \textit{FLOP/s limit compliance.} Ensuring the data center's total computing power is below agreed thresholds.
    \item \textit{Certified chip usage.} Verifying that only approved chip models are in use.
    \item \textit{Security measures.} Verifying implementation of required security protocols.
    \item \textit{Training run evidence.} Examining records and transcripts of large-scale AI training activities.
    \item \textit{Hardware integrity.} Inspecting for any evidence of chip tampering \citep{Aarne2024}.
\end{itemize}

In addition to requiring periodic inspections, an agreement could also require \textit{continuous monitoring} of certain facilities. In a continuous monitoring setup, inspectors are present at facilities at all times to catch any violations of agreements (such as tampering with hardware). A final possible implementation is challenge inspections, similar to those conducted by the Organization for the Prohibition of Chemical Weapons (OPCW), where inspections can be called for on short notice based on suspicions of non-compliance \citep{CWC1997}. 

\textbf{Precedent.} The New START treaty signed by the USA and Russia provides for 18 annual on-site inspections for the American and Russian inspections (US Department of State). These inspections allow for specific verification activities, such as confirming the number of reentry vehicles on deployed missiles, counting nuclear weapons on bombers, and verifying the conversion or elimination of weapon systems. The treaty's approach of allowing a limited number of highly structured inspections, focused on counting and verifying specific hardware, is a suggestive precedent for inspections of data centers. Notably, an earlier treaty (START I) also provided for continuous monitoring of specific facilities \citep{arms2022start}.

The most significant precedent for the detailed inspection of hardware is the IAEA’s mandated use of bespoke tamper-evident containment seals for nuclear materials \citep{IAEA2011}. These seals – each of which bears a unique identifier – are designed to provide clear evidence of any tampering or unauthorized access. IAEA inspectors examine these seals during on-site visits, allowing them to detect any undeclared movement or use of nuclear materials.  

\subsubsection{ON-SITE INSPECTIONS OF SEMICONDUCTOR MANUFACTURING FACILITIES}

Inspections of semiconductor manufacturing facilities could be used to determine the quantity and nature of chips produced. The manufacturing of advanced chips is a highly specialized activity, and only a few entities have this capacity \citep{Sastry2024}. For example, it is well known that ASML produces EUV lithography systems which are needed to manufacture the latest generation of advanced chips \citep{khan2021semiconductor}. If inspectors identified the existence of such machines, it would be relatively easy to know what kind of chips are possible to construct at the manufacturing site. Inspectors may also be able to use basic metrics like the square-meterage of a facility or the number of lithography machines to bound the number of chips that are possible to produce in such facilities.

These inspections could also verify that facilities are producing chips in accordance with any hardware-related agreements. For example, if nations agreed to only build chips with certain on-chip hardware governance mechanisms, inspections of semiconductor manufacturing facilities could identify non-compliance. Inspectors could look at a sample of chips, potentially midway through production, to ensure they have the correct mechanisms. 

As with the inspections of data centers, continuous monitoring could also be used for semiconductor manufacturing facilities. 

\textbf{Precedent.} The use of on-site inspections for monitoring compliance with international agreements has been well-established in other domains, particularly in controlling extreme risks.

\begin{enumerate}
    \item \textit{Organization for the Prohibition of Chemical Weapons (OPCW).} The OPCW conducts inspections at facilities that produce toxic chemicals and their precursors. These inspections involve an initial tour, followed by a detailed inspection plan, physical inspections, and a review of the facility's records to verify compliance. The intensity and duration of inspections vary depending on the perceived risk, with chemicals categorized into three schedules based on their threat level \citep{OPCW2024}.
    
    \item \textit{Preparatory Commission for the Comprehensive Nuclear-Test-Ban Treaty Organization (CTBTO).} The CTBTO, although not fully operational due to the Comprehensive Nuclear-Test-Ban Treaty's pending entry into force, has established protocols for on-site inspections (OSI). These inspections are intended to verify compliance with the treaty, particularly in detecting and investigating potential nuclear explosions. If the treaty enters into force, an OSI could be initiated upon the request of a State Party. The inspection area could cover up to 1000 km$^2$ \citep{CTBTO2024}.
\end{enumerate}

\subsubsection{ON-SITE INSPECTIONS OF AI DEVELOPERS}

An international inspection team could visit an AI development facility to ensure that developers are running authorized code, properly implementing model evaluations and safeguards, and assess safety culture and security concerns. Inspections could involve various components, such as reviewing code \citep{auditing-black-box-is-not-good}, assessing compliance with commitments from safety cases\footnote{For example, whether mandated interpretability techniques are implemented \citep{transparent-ai-survey,bereska2024mechanisticinterpretabilityaisafety}, or evaluations \citep{shevlane2023modelevaluationextremerisks, phuong2024evaluatingfrontiermodelsdangerous,anderljung2023publiclyaccountablefrontierllms}.}, and conducting semi-structured interviews with key personnel to solicit security-relevant concerns (see \citet{wasil2024understanding}). Inspections could uncover the usage of unauthorized or unlicensed AI algorithms. A number of privacy-preserving technologies in development could facilitate such inspections without being overly intrusive. .\footnote{We thank Mauricio Baker for this suggestion, which is also included in his upcoming report.}  

\textbf{Precedent.} 
The closest precedent is the IAEA’s use of on-site inspections, as discussed above. Their approach demonstrates the feasibility of conducting thorough on-site inspections in sensitive, high-tech environments, which could be adapted for AI development facilities. The key difference is that AI inspections would focus more on software and computational resources rather than physical materials, requiring inspectors with specialized expertise in AI technologies and development practices. 

\begin{figure}[ht!]
    \centering
    \includegraphics[width=1\linewidth]{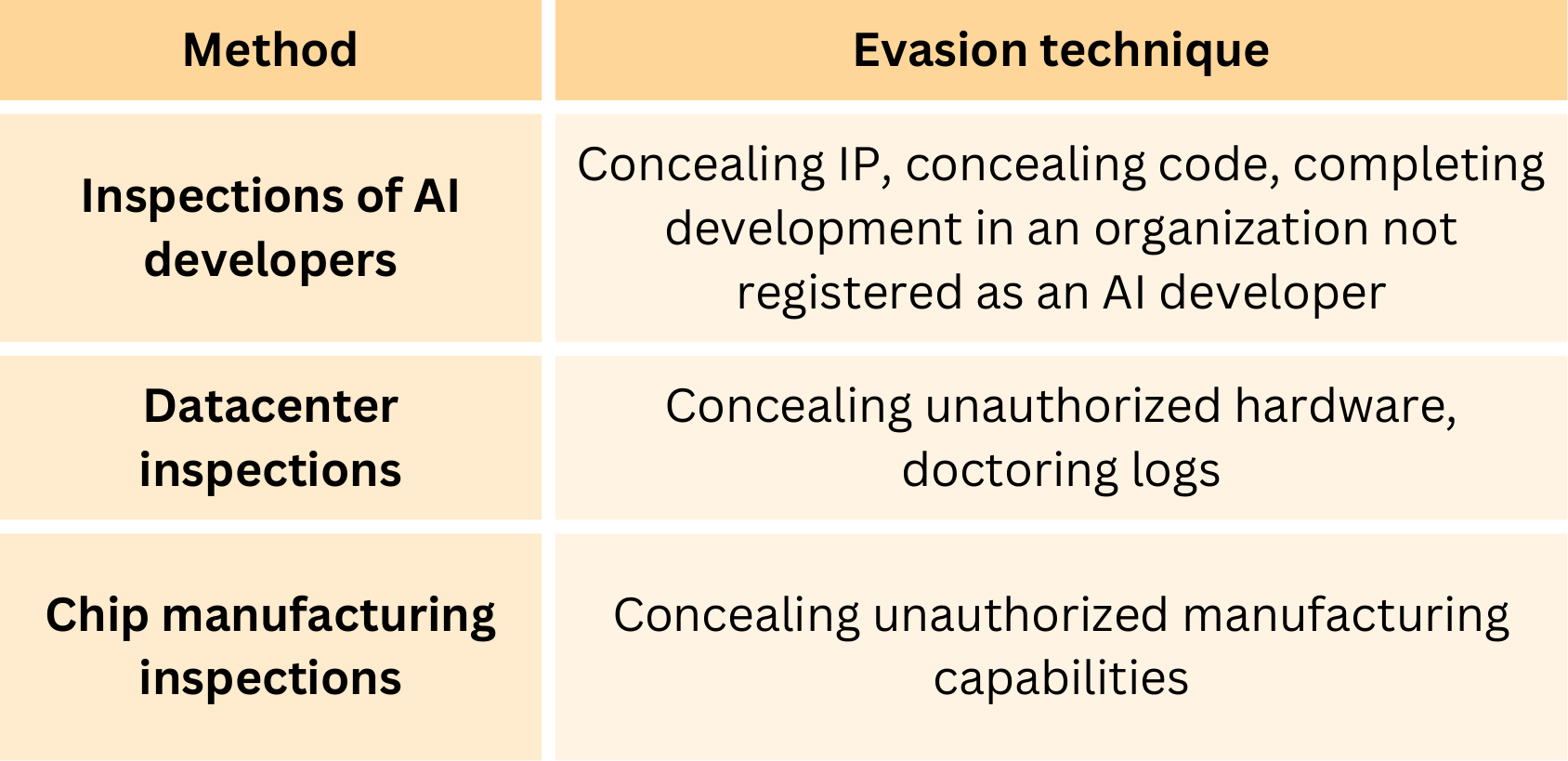}
    \caption{Summary of evasion techniques to avoid access-dependent verification methods}
    \label{fig:access-dependent-counters}
\end{figure}

\subsubsection{KEY TAKEAWAYS}
Access-dependent methods can allow for in-depth inspections of key facilities such as AI development facilities, hardware manufacturing facilities, and data centers. If international inspectors have sufficient access to these facilities, this provides a great deal of robustness to a verification regime. However, such methods may be perceived as invasive, and they may rely on the permission of nations that are suspected of unauthorized activity. Access-dependent methods can also be somewhat flexible depending on the amount of political will and the level of access that nations are willing to provide. To preserve privacy or trade secrets, inspectors may receive limited access– enough access to verify that an unauthorized training run is not being conducted but not enough access to see exactly what kind of tasks are being performed. 

\subsection{Hardware-dependent verification methods}
\label{sec:hardware-dependent-verif}

\subsubsection{CHIP LOCATION TRACKING}

Chip location tracking involves implementing a system to monitor the movement and use of AI-capable chips \citep{Brass2024}. This method requires international agreement on chip manufacturing standards and the implementation of tracking mechanisms directly into the hardware. Each chip above a certain computational threshold would be assigned a unique identifier and equipped with secure tracking capabilities.

\textbf{Precedent.} The concept of tracking and monitoring critical technology has several precedents across different industries, particularly where security, compliance, and international regulation are concerned.

\begin{itemize}
    \item \textit{Nuclear Material Tracking.} The IAEA monitors and tracks nuclear materials globally using systems like the Integrated Nuclear Fuel Cycle Information System \citep{IAEA-NFCIS}. The IAEA also maintains a databse of incidents involving trafficking or other unauthorized uses of nuclear or radioactive materials \citep{IAEA-ITDB}.
    
    \item \textit{Pharmaceutical Supply Chain Tracking.} The Drug Supply Chain Security Act (DSCSA) in the United States currently outlines steps to achieve, ``an interoperable and electronic way to identify and trace certain prescription drugs at the package level as they move through the supply chain'' \citep{FDA-DSCSA}.
\end{itemize}

\subsubsection{CHIP-BASED REPORTING}

Chip-based reporting involves implementing  mechanisms within AI-capable chips and closely associated hardware (e.g., networking cards) to automatically detect and signal when they are being used in ways that violate agreed-upon constraints. These constraints might include thresholds on the number of chips connected together, or specific operations the chip is not authorized to perform. By embedding these reporting mechanisms at the lowest levels of the software stack --- within the firmware and drivers of the AI-capable chips or associated networking devices --- it may become more challenging for developers to bypass these safeguards. As one moves up the software stack, toward components that operate at higher levels of abstraction, it becomes easier for developers to replace authorized programs with their own software, potentially circumventing the constraints. Therefore, focusing on the lower levels of the stack, such as firmware, which is the (often read-only) software residing on the device \citep{nasa2004software}, and the drivers, which allow the operating system to communicate with the device \citep{microsoft2023driver}, is crucial for effective enforcement of constraints. These components are typically developed by the chip maker, further limiting the number of developers who could foreseeably edit reporting mechanisms.

\textbf{Precedent.}
The closest precedent for this type of firmware-based reporting is the Light Hash Rate (LHR) GPUs developed by NVIDIA. These GPUs can detect, via mechanisms implemented in their firmware and drivers, whether they are being used for Ethereum mining \citep{nvidia2021geforce}. Similar strategies could foreseeably be developed to report unauthorized AI training.

\begin{figure}[ht!]
    \centering
    \includegraphics[width=1\linewidth]{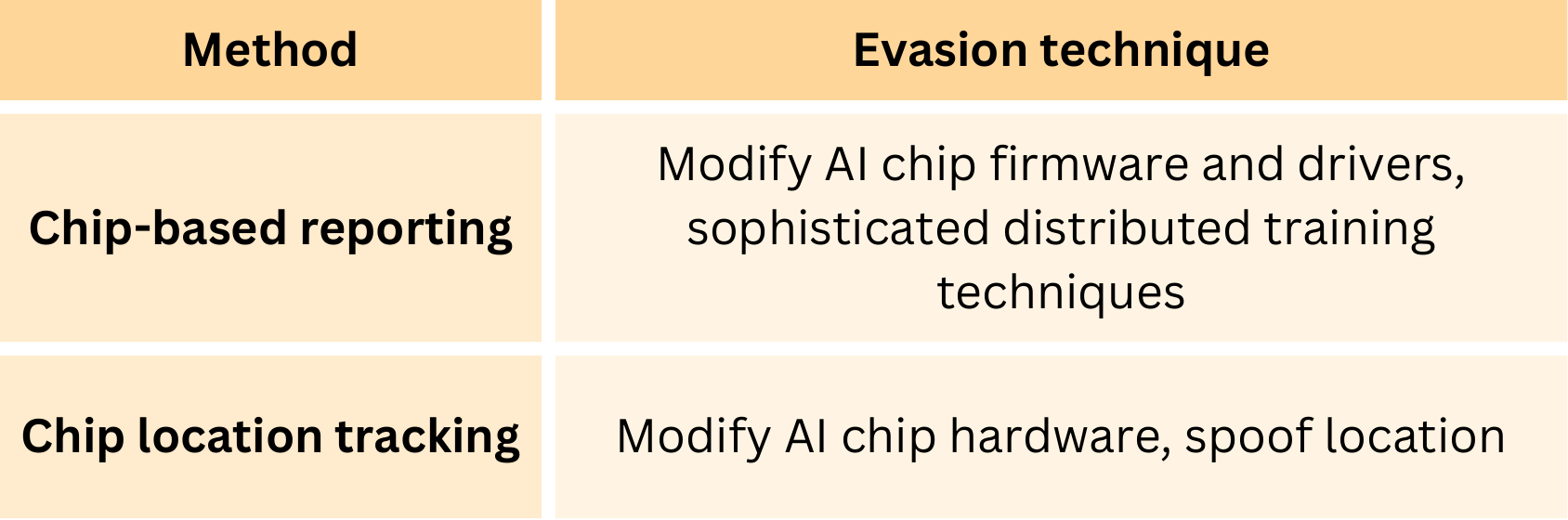}
    \caption{Summary of evasion techniques to avoid hardware-dependent verification methods.}
    \label{fig:hardware-dependent-counter}
\end{figure}

\subsubsection{KEY TAKEAWAYS}

Hardware-dependent verification methods may offer robust and privacy-preserving tools for detecting non-compliance. However, these methods require nations with advanced hardware manufacturing capabilities to agree to rules around hardware manufacturing. Another challenge is that advanced chips are already in circulation (without hardware-enabled mechanisms built-in). A verification regime relying on hardware-dependent measures may need to address this ``legacy hardware", potentially through retrofitting techniques or gradual phase-out strategies.

If successfully implemented, these methods could dramatically enhance the effectiveness of other verification approaches, particularly on-site inspections. However, they also raise important concerns about privacy, national sovereignty, and potential misuse that must be carefully addressed. Overall, hardware-dependent methods represent a promising but long-term goal, requiring sustained international cooperation and technological innovation to realize their full potential in AI governance.

\section{Limitations and discussion}

This paper examined verification methods that could help nations detect non-compliance with international agreements prohibiting unauthorized AI development and unauthorized data centers.

Verification methods vary in their feasibility, intrusiveness and effectiveness. National technical means offer a valuable starting point, capable of detecting large-scale AI infrastructure and unusual patterns in energy consumption, hardware imports, and financial transactions. However, national technical means are limited in their ability to identify software-level violations or concerted attempts at concealment. Access-dependent methods, such as on-site inspections, provide more robust reassurance but require nations to agree to international inspections. Hardware-dependent approaches offer additional robustness (potentially even guarantees) but face some implementation challenges, including the need to address existing legacy hardware.

\textbf{Table 1} summarizes gaps in individual verification methods, as well ways each method can be complemented by other methods.

Additionally, the verification methods are at different levels of maturity: some are ready-to-implement, while others will require additional research. \textbf{Figure 5} lists the verification methods based on the amount of additional research required to implement each method. 

\section{Future research directions}

Our work provides a starting point for discussions about verification methods, but there are many open questions that can be addressed by future work. Some of these directions include:

\begin{itemize}
    \item \textbf{Red-teaming exercises for international verification}. In a ``red-team" step, the authors could brainstorm how an adversary might try to hide an unauthorized training run or unauthorized data center. Then, in a ``blue team" step, the authors could identify how one or more verification methods could catch the adversary. Then, in a subsequent ``red team" step, the authors could brainstorm if there are feasible ways for the adversary to avoid or undermine the verification method(s). This process could be used to determine likely ways that adversaries may try to evade verification methods and highlight ways of strengthening international verification regimes.
    
    \item \textbf{Design of international AI governance institutions.} Compliance with international agreements is often verified by international institutions. Some early work has proposed international organizations that could set and verify compliance with safety standards \citep{ho2023international, cigiframework}, certify national licensing agencies \citep{trager2023international}, verify compliance with a variety of potential agreements (see \citet{maas2023international}), and participate in joint AI safety research \citep{cigiframework}. One avenue for future research is to provide more details about how an international verification agency could be structured, how decision-making power is distributed between nations, how the agency handles disputes over non-compliance, and what powers ought to be granted to the agency. Such work could draw from best practices or lessons learned from the design and implementation of other international institutions (such as the IAEA and the OPCW) and bilateral or multilateral agreements (such as the Strategic Arms Reduction Treaties and the Wassenaar Agreement).
    
    \item \textbf{Enforcement of international agreements.} Our paper focused on \textit{verification}-- detecting whether or not nations are complying with an agreement. A separate important question is \textit{enforcement}-- how nations should react in the event that non-compliance is identified. Such work could examine what kinds of responses would be proportionate to the violation. For example, evidence of small-scale chip smuggling would warrant a less strong response than evidence of an illegal or unauthorized training run.
    
    \item \textbf{Research on hardware-enabled mechanisms to enhance verification and/or enforcement.} Hardware-enabled mechanisms can unlock new verification methods and make existing verification methods more robust. Some hardware-enabled mechanisms are ready to be implemented swiftly, while others may take several years of research to further develop. Additionally, there are open questions relating to how to make hardware-enabled mechanisms more tamper-proof and privacy-preserving (see \citet{kulp2024hardware}).
    
    \item \textbf{Detecting unauthorized AI deployment or inference.} Our paper focuses on detecting unauthorized AI \textit{development}. Nations may also wish to have agreements in which they agree not to \textit{deploy} advanced AI systems in certain ways (for example, nations might prohibit AI from being deployed in the context of nuclear systems, military R\&D research, or AI R\&D research that could trigger uncontrolled AI development.) Future work could examine verification methods that could detect the unauthorized deployment of AI systems, potentially through hardware-enabled licenses that detect the presence of unauthorized code used for inference.
    
    \item \textbf{Detecting compliance with agreements around model evaluations}. International agreements may require that certain kinds of model evaluations are conducted to detect potential safety or security issues (see \citet{shevlane2023modelevaluationextremerisks}). Reliable risk evaluations and risk mitigation strategies could become a minimum safety bar imposed by international agreements. Future work could examine verification methods that allow international authorities to ensure that parties are implementing a set of internationally-required model evaluations, as well as any specific model evaluations that a developer proposed as part of a safety case or licensing application (see \cite{clymer2024safety,wasil2024affirmative}).
    
    \item \textbf{Actions the international community can take in the immediate future.} In the future, nations may be concerned enough about AI global security risks to warrant ambitious international agreements that require verification methods. For the immediate future, however, nations are interested in improving their understanding of global security risks. There are many actions that governments and civil society groups can participate in to increase global understanding of AI progress and AI risks. Examples include efforts like the UK and Seoul AI Safety Summits (see \citet{bletchley2023declaration}), the establishment of the US and UK AI Safety Institutes and the Chinese AI Safety Network, Track II Dialogues between Western scientists and Chinese scientists (see \citet{idais2023}), and plans for how to respond to AI-related emergencies (see \citet{wasil2024aiprep}).
\end{itemize} 

\section{Conclusion}
Our work provides an initial step toward a better understanding of how compliance with international AI agreements could be verified. Efforts to improve our understanding of verification methods will be especially important if global security risks from advanced AI become concerning enough to motivate coordinated national and international action. We believe some AI governance work should aim to prepare in advance for such scenarios. Such ``future-oriented" AI governance work could address questions that would inform policymaking efforts in scenarios where concerns about global security risks became significantly stronger. Our hope is that our work on verification methods illustrates an example of promising work in this category. 

\bibliography{aaai25}

\onecolumn

\appendix

\section{Limitations of methods and possible solutions using complementary methods}

\begin{longtable}{|p{3cm}|p{5cm}|p{6cm}|}
    \caption{Limitations of methods and possible solutions using complementary methods.}
    \label{tab:lim-and-gaps} \\
    \hline
    Verification method & Primary limitations & Complementary methods \\
    \hline
    \endfirsthead

    \hline
    Verification method & Primary limitations & Complementary methods \\
    \hline
    \endhead

    \hline
    \endfoot

    \hline
    \endlastfoot

    Satellite imagery & Data centers could be concealed underground or camouflaged & \textit{National intelligence services} can provide human intelligence and signals intelligence to identify hidden facilities that satellite imagery might miss. They can gather information on construction activities, personnel movements, and communications that could indicate the presence of a concealed data center.\vspace{0.5cm}
        
    \textit{Energy monitoring} complements satellite imagery by detecting unusual power consumption patterns that might indicate a hidden data center. Even if a facility is visually concealed, its energy requirements are difficult to hide, especially for large-scale AI operations.\vspace{0.5cm}

    \textit{Chip location tracking} can determine the approximate location of data centers and discourage concealment. \\
    \hline
    Whistleblowers & Reliability issues: Whistleblowers may provide incomplete, biased, or false information.\vspace{0.5cm}
        
    Limited access: Not all potential whistleblowers have access to critical information.\vspace{0.5cm}

    Fear of retaliation: Potential whistleblowers may be deterred by fears of personal or professional consequences.
     & \textit{National intelligence services} can corroborate or refute whistleblower claims through other intelligence gathering methods.\vspace{0.5cm}
 
    \textit{On-site inspections} can be triggered by whistleblower reports, allowing for direct verification of claims. Inspectors can look for specific evidence pointed out by whistleblowers, increasing the effectiveness of the inspection.\vspace{0.5cm}

    Financial intelligence can be used to verify claims about resource allocation or unusual transactions mentioned by whistleblowers. \\
    \hline
    National intelligence services & Using national intelligence can be unnecessarily invasive and infringe on national sovereignty. The source of intelligence is often classified, which makes it a poor foundation for transparent discussion in international forums. & \textit{Satellite imagery} can provide visual confirmation of intelligence reports about suspected facilities, offering a less intrusive method of verification.\vspace{0.5cm} 

    \textit{Financial intelligence} can corroborate intelligence about resource allocation and unusual transactions, providing a paper trail for activities identified through other intelligence means.
    \\
    \hline
    Energy monitoring & Though plausible in theory, this method is unproven in practice. Energy consumption may be disguised as other high-energy activities. Obtaining detailed energy consumption data is also likely to be challenging. & Customs data analysis can corroborate energy monitoring data by tracking the import of high-performance computing equipment to areas with suspicious energy consumption patterns.
    \\
    \hline
    Customs data & Countries with advanced domestic manufacturing capabilities may be able to produce key components internally, reducing the effectiveness of customs data analysis. Distinguishing between components intended for authorised and unauthorised is likely to be challenging. &\textit{On-site inspections of semiconductor manufacturing} facilities can verify whether domestic production capabilities match declared capacities, helping to identify discrepancies that might indicate undeclared production bypassing customs.\vspace{0.5cm}

\textit{Chip location tracking}, if implemented, can help verify the end destination and use of key components that have passed through customs, ensuring they are being used as declared.
    \\
    \hline
    Financial intelligence & Many AI-related purchases may have legitimate alternate uses, making it difficult to distinguish between authorized and unauthorized activities. This method may be disproportionately invasive. 
Financial intelligence is also limited by banking secrecy laws and a potential lack of international cooperation.
 &\textit{Customs data analysis} can corroborate financial intelligence by providing physical evidence of hardware purchases and movements that correspond to suspicious financial transactions.\vspace{0.5cm}

\textit{Whistleblowers} can provide insider information about financial practices, helping to interpret complex transactions or reveal hidden financial structures used to fund unauthorized AI development.
    \\
    \hline
    Data center inspections & Inspections can only be carried out with the agreement of the host nation, potentially allowing time for concealment of violations. Thorough inspections are also both invasive and very resource-intensive, requiring significant time, expertise and resources. 
 &\textit{Chip location tracking}, if implemented, can be verified during inspections to ensure that the physical location of AI-capable chips matches their reported locations.\vspace{0.5cm}

\textit{Whistleblower} information can guide inspectors to look for specific evidence of non-compliance that might otherwise be overlooked.
    \\
    \hline
    Fab inspection & Like all inspections, these inspections are also resource-intensive, invasive and pose threats to intellectual property. The technological complexity of chip manufacturing may also make it challenging for inspectors to detect potential violations without highly specialized expertise.
 &\textit{Chip location tracking}, if implemented, can be initiated during the inspection process, ensuring that newly manufactured AI-capable chips are properly registered and tracked from the point of production.
    \\
    \hline
    AI developer inspection & Unlike hardware, software can be quickly modified or hidden, making violations difficult to detect. Such inspections also require highly specialized knowledge, and may pose a disproportionate risk to proprietary algorithms and research. 
 &\textit{Whistleblowers} can provide insider information about development practices, guiding inspectors to specific areas or systems of concern.\vspace{0.5cm}

\textit{Financial intelligence} can be cross-referenced to ensure declared AI projects match financial records.
    \\
    \hline
    Chip location tracking & Requires agreement on chip manufacturing standards. Sophisticated actors may find ways to disable tracking mechanisms. The effectiveness of this intervention would be limited to the production of new chips. 
 &\textit{On-site inspections}of manufacturing sites can ensure that chips are being made with the required location tracking mechanisms.\vspace{0.5cm}

\textit{Satellite imagery} can provide additional, more precise location tracking.
    \\
    \hline
    Fixed set reporting & Requires agreement on chip manufacturing standards. False positives/negatives: Balancing sensitivity to catch violations without triggering false alarms is difficult.
The effectiveness of this intervention would be limited to the production of new chips. 
 &\textit{On-site inspections} of data centers can be triggered by automatic signals of non-compliance, allowing for rapid verification of potential violations.
    \\
    \hline
    Firmware-based reporting & Requires agreement on chip manufacturing and implementation standards; difficult to implement; may come with an economic or computational cost.
 &\textit{On-site inspections} could be triggered by automatic signals of non-compliance, allowing for rapid verification of potential violations.
    \\
    \hline
\end{longtable}

\newpage

\section{Research and development estimates for verification methods}

\begin{figure}[ht!]
    \centering
    \includegraphics[width=0.7\linewidth]{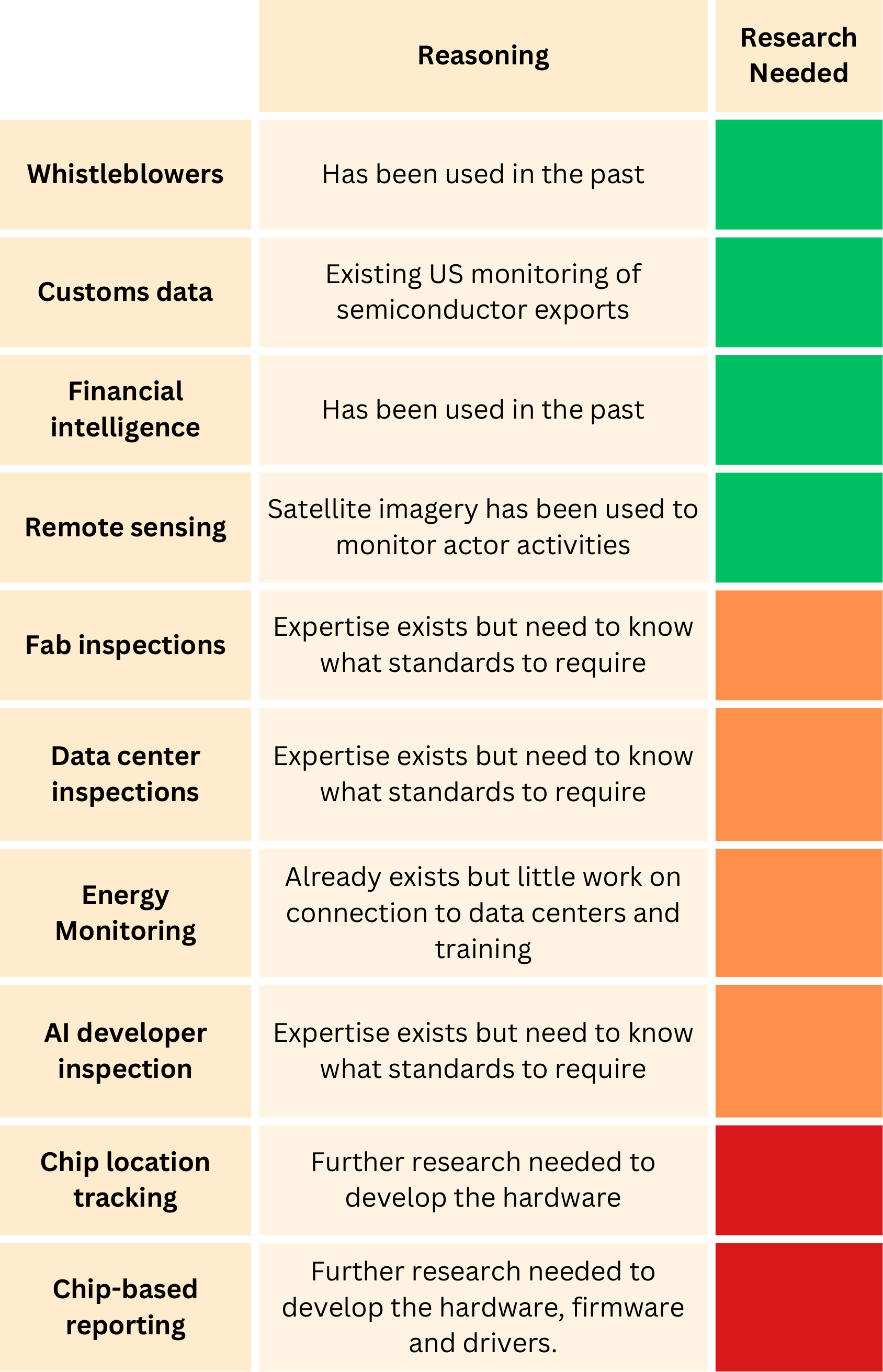}
    \caption{Estimated research and development needed for verification methods investigated. Note that green indicates little additional research needed, orange indicates some additional research and red indicates significant additional research.}
    \label{fig:research-est}
\end{figure}

\newpage

\section{List of verification methods for international AI agreements}

\begin{figure}[ht!]
    \centering
    \includegraphics[width=0.7\linewidth]{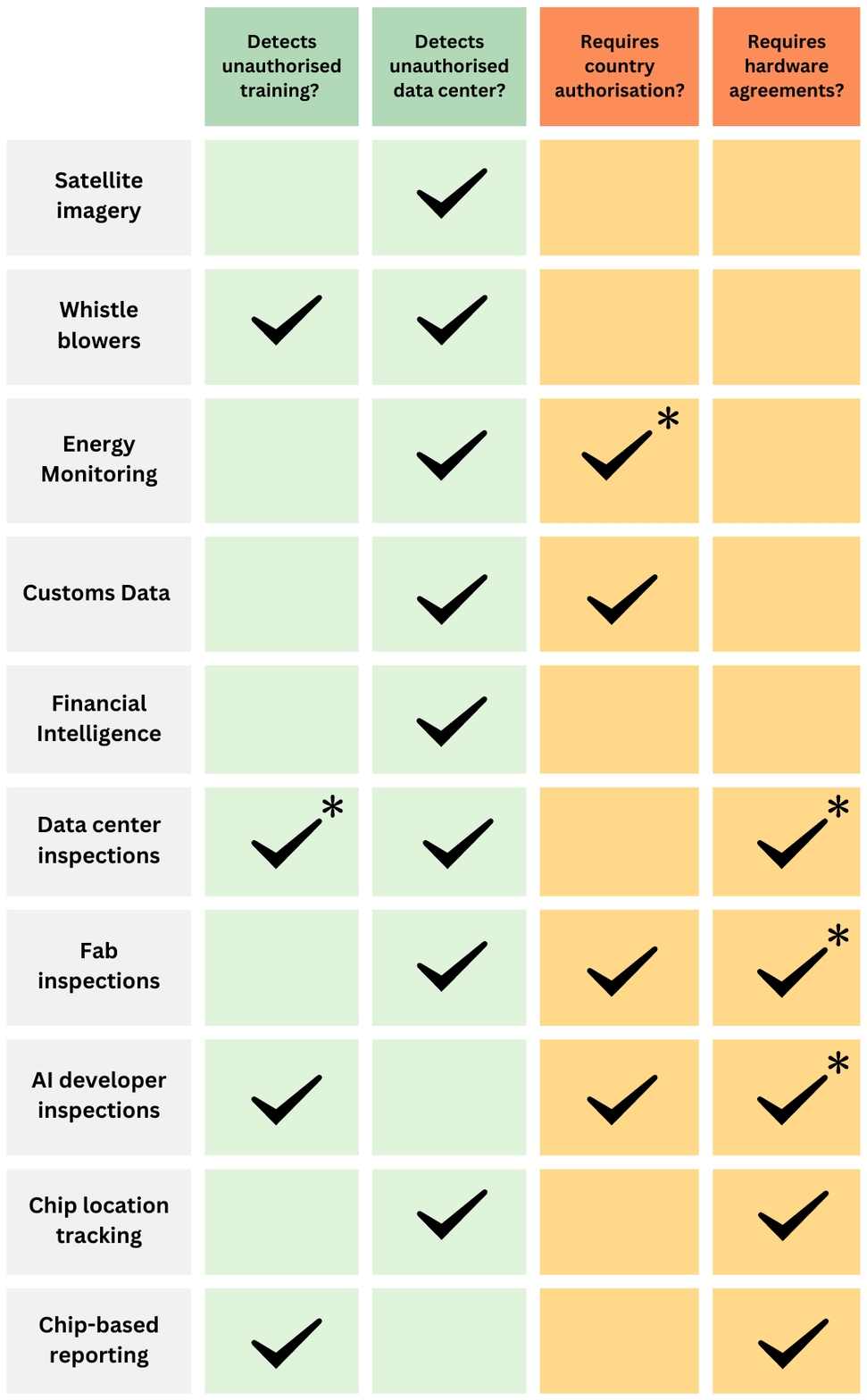}
    \caption{\small We identify 12 verification methods (left-hand column) which could be used to investigate compliance with international agreements\textbf{.} Each method is categorized by: (a) whether or not it can be used to detect unauthorized training runs, (b) whether or not it can be used to detect unauthorized data centers, (c) whether or not the use of the method requires authorization from the suspected entity, and (d) whether or not the method relies on agreements relating to the development or distribution of advanced hardware. Asterisks (*) indicate nuanced applications. \textbf{Energy monitoring:} it may be possible to infer unauthorized training by detecting energy consumption patterns at known data centers that exceed those suggested by reported levels. \textbf{Data center inspections:} While basic inspections can be conducted without agreements mandating the use of specific hardware, the effectiveness is significantly enhanced by such agreements (hardware-enabled chip logs). \textbf{Fab inspections:} while basic inspections are useful without agreements, their relevance is greatest as a tool to verify that labs are producing hardware compliant with mandated standards. \textbf{National intelligence services:} While these services may operate without explicit authorization from the suspected entity, their use in verification contexts would ideally be governed by international agreements to ensure legitimacy and prevent potential diplomatic conflicts.}
    \label{fig:ast-table}
\end{figure}

\end{document}